\documentclass[aps,prl,twocolumn,showpacs,superscriptaddress]{revtex4-1}
\usepackage{graphicx}
\usepackage{dcolumn}
\usepackage{bm}
\usepackage{xcolor}

\usepackage[utf8]{inputenc}
\usepackage[T1]{fontenc}
\usepackage{mathptmx}
\usepackage{etoolbox}
\usepackage{amsfonts}
\usepackage{amsmath}
\usepackage{physics}
\usepackage{enumerate}
\usepackage[normalem]{ulem}

\begin{document}

\title{Entanglement and Quantum Correlation Measures from a Minimum Distance Principle}

\author{Arthur Vesperini}
\affiliation{DSFTA, University of Siena, Via Roma 56, 53100 Siena, Italy}
\affiliation{QSTAR \& CNR - Istituto Nazionale di Ottica,    Largo Enrico Fermi 2, I-50125 Firenze, Italy}
\affiliation{INFN Sezione di Perugia, I-06123 Perugia, Italy}

\author{Ghofrane Bel-Hadj-Aissa}
\affiliation{DSFTA, University of Siena, Via Roma 56, 53100 Siena, Italy}
\affiliation{QSTAR \& CNR - Istituto Nazionale di Ottica,    Largo Enrico Fermi 2, I-50125 Firenze, Italy}
\affiliation{INFN Sezione di Perugia, I-06123 Perugia, Italy}

\author{Roberto Franzosi}
\email[]{roberto.franzosi@unisi.it}
\affiliation{DSFTA, University of Siena, Via Roma 56, 53100 Siena, Italy}
\affiliation{QSTAR \& CNR - Istituto Nazionale di Ottica,    Largo Enrico Fermi 2, I-50125 Firenze, Italy}
\affiliation{INFN Sezione di Perugia, I-06123 Perugia, Italy}

\date{\today}

\begin{abstract}
Entanglement, and quantum correlation, are precious resources for quantum technologies implementation based on quantum information science, such as, for instance, quantum communication, quantum computing, and quantum interferometry.
Nevertheless, to our best knowledge, a directly computable measure for the entanglement of multipartite mixed-states is still lacking. 
In this work, we derive, from a minimum distance principle, an explicit measure of the degree of quantum correlation for pure or mixed multipartite states; we further propose an entanglement measure, derived  from the quantum correlation measure using a novel regularization procedure for the density matrix. 
Then, a comparison of the proposed measures, of quantum correlation and entanglement, allows one to distinguish between quantum correlation detached from entanglement and the one induced by entanglement, hence to define the set of separable but non-classical states. 

Since all the relevant quantities in our approach, descend from the geometric structure of the projective Hilbert space, the proposed method is of general application. 

Finally, we apply the derived measures as an example to a general Bell diagonal state and to the Werner states, for which our regularization procedure is easily tractable.

\end{abstract}

\maketitle

\section{\label{sec:intro} Introduction}

Entanglement has assumed an important role in quantum information theory and in the development of the quantum technologies. It is considered as a valuable resource in quantum cryptography, in quantum computation, in teleportation and in quantum metrology \cite{GUHNE20091}. Nevertheless, entanglement remains elusive and the problem
of its quantification in the case of a general system, is still open
\citep{PhysRevA.95.062116,PhysRevA.67.022320}.
In the last decades, several approaches have been developed to quantify the entanglement in the variety of states of the quantum realm. However, the rigorous achievements in the explicit quantification of entanglement, are limited to bipartite systems case \cite{RevModPhys.81.865}.
Entropy of entanglement is uniquely accepted as measure of entanglement for pure states of bipartite systems \cite{PhysRevA.56.R3319}, while entanglement of formation \cite{PhysRevLett.80.2245}, entanglement distillation \cite{PhysRevA.54.3824,PhysRevLett.76.722,PhysRevLett.80.5239} and relative entropy of entanglement \cite{PhysRevLett.78.2275} are largely acknowledged as faithful measures  for bipartite mixed systems \cite{Adesso_2016}.
An extensive literature is devoted to the study of entanglement in multipartite systems. Over time, different approaches have been proposed including, e.g. in the case of pure states, the study of the equivalence classes in the set of multipartite entangled states \cite{PhysRevA.62.062314,briegel_PRL86_910},  whereas, the study of entanglement in mixed multipartite states have been addressed, e.g., with a Schmidt measure \cite{PhysRevA.64.022306} or with a generalisation of concurrence \cite{PhysRevA.61.052306,PhysRevLett.93.230501}.
In the last years, have been proposed entanglement estimation-oriented approaches and derived from a statistical distance \cite{PhysRevLett.72.3439} concept, as, for instace, the quantum Fisher information \cite{PhysRevLett.102.100401,PhysRevA.85.022321,PhysRevA.85.022322,j.aop.2019.167995}.
 
In a recent paper \cite{PhysRevA.101.042129}, some of us have put forward a new entanglement measure able to quantify by a direct calculation, the degree of entanglement of a general multi-qudit hybrid pure state. Such entanglement measure, named entanglement distance (ED), derives from a minimum distance principle. In Ref. \cite{PhysRevA.101.042129}, we have shown that a pure state is entangled if manifests any quantum correlation between its components and vice versa. In fact, entanglement and correlation are completely equivalent in the case of pure states . On the contrary, in the case of mixed states, one can observe states that manifest correlations detached from entanglement. In the present work, we propose a new measure of quantum correlation, and a related entanglement measure for mixed states, derived from the first through a regularization process.
Such entanglement measure derives from a minimum distance principle and, for this reason, we name it entanglement distance.

In this article we first summarize the ED derivation in the case of pure states, then we derive a quantum correlation measure for mixed states from a minimum distance principle and, from the latter, we derive an entanglement measure valid for the same class of states.
Finally, we report as easily tractable examples the application of the quantum correlation measure and ED to a general Bell diagonal state and to the Werner states.

\section{\label{sec:pure} Entanglement Distance for Pure States}

The ED \cite{PhysRevA.101.042129} derives from the Fubini-Study metric which gives the distance between two neighbouring states of a finite projective Hilbert-space \cite{gibbons} according to
\begin{equation}
d^2_{_{FS}} (|s\rangle,|s\rangle +|ds\rangle)=
\langle d s | d s \rangle - \dfrac{1}{4}
|\langle s | d s \rangle - \langle d s | s \rangle|^2 \, ,
\label{F-S-metric}
\end{equation}
$|s \rangle$ is a normalized state and $|ds \rangle$
is an infinitesimal variation of such state. We consider the Hilbert space
${\cal H } = {\cal H}^0 \otimes {\cal H}^1 \cdots {\cal H}^{M-1}$ tensor product of $M$ two qubits Hilbert spaces.
The entanglement is invariant under local unitary transformations, thus a measure quantifying it must be a function of the set of classes of equivalence of states that differ for local unitary transformations. To derive a measure with this property from $d_{_{FS}}$, for any state $|s\rangle$, we first consider the set of states 
\begin{equation}
|U,s\rangle = \prod^{M-1}_{\mu=0} U^\mu |s\rangle \, ,
\label{Us}
\end{equation}
where, for $\mu=0,\ldots,M-1$, $U^\mu$ is an arbitrary $SU(2)$ (local) unitary operator that operates on the $\mu$th qubit. The states \eqref{Us} share the same degree of entanglement.
We consider an infinitesimal variation of state \eqref{Us}, given by
\begin{equation}
|dU,s\rangle = \sum^{M-1}_{\mu=0} d\tilde{U}^\mu |U,s\rangle \, ,
\label{dUs}
\end{equation}
where 
\begin{equation}
d\tilde{U}^\mu = -i 
( \bm{\sigma}_{\bf n})^\mu
d \xi^\mu 
\label{dU}
\end{equation}
rotates the $\mu$th qubit by an infinitesimal angle $2 d\xi^\mu$  around
the unitary vector ${\bf n}^\mu $. Here and in the following we use the notation $(\bm{\sigma}_{\bf n})^\mu = ({\bf n}^\mu \cdot \bm{\sigma}^\mu)$, and 
for $\mu=0,\ldots,M-1$, we denote by $\sigma^\mu_1$, $\sigma^\mu_2$ and $\sigma^\mu_3$
the three Pauli matrices operating on the $\mu$-th qubit, where the index $\mu$ labels the spins.
By substitunig $|U,s\rangle$ and $|dU,s\rangle $ in Eq. \eqref{F-S-metric}, in place of $|s\rangle$ and $|ds\rangle$, respectively, we get 
\begin{equation}
    d^2_{_{FS}} (|U,s\rangle,|U,s\rangle +|dU,s\rangle) =
    \sum_{\mu \nu }g_{\mu \nu} (|s\rangle,{\bf v}) d\xi^\mu d\xi^\nu \, ,
\end{equation}
where, the Fubini-Study tensor metric is
\begin{equation}
    g_{\mu \nu} (|s\rangle,{\bf v}) =
\langle s | (\bm{\sigma}_{\bf v})^\mu
(\bm{\sigma}_{\bf v})^\nu|s\rangle 
-\langle s | (\bm{\sigma}_{\bf v})^\mu
|s\rangle
\langle s |
(\bm{\sigma}_{\bf v})^\nu|s\rangle \, ,
\label{gmunu}
\end{equation}
${\bf v} = ({\bf v}^0,\ldots,{\bf v}^{M-1})$ and
the unit vectors ${\bf v}^\mu$, $\mu=0,\dots,M-1$, are derived by a rotation of the original
ones of Eq. \eqref{dU}, according to ${\bf v}^\nu \cdot \bm{\sigma}^\nu =
U^{\nu \dagger} {\bf n}^\nu \cdot \bm{\sigma}^\nu
U^{\nu}$,
where there is no summation on the index $\nu$.
Of course, for each state $|s\rangle$, the tensor metric $g_{\mu \nu} (|s\rangle,{\bf v})$ is not invariant under rotation of the unit vectors ${\bf v}^\mu$. In order to derive a measure that is invariant, we define the entanglement measure of $|s\rangle$, as the inferior value of the trace of $g_{\mu \nu} (|s\rangle,{\bf v})$ as the orientation of the unit vectors ${\bf v}^\mu$ change, in formulas
\begin{equation}
E(|s\rangle) =\inf_{ \{{\bf v}^\nu\}_\nu} \tr (g(|s\rangle,{\bf v})) \, ,
\label{emeasure}
\end{equation}
where $\tr$ is the trace operator and where the $\inf$ is taken over all possible orientations of the unit vectors $ {\bf v}^\nu$ ($\nu = 0, \dots , M-1$).
We emphasize that, in general, the inspection of the block structure of $g(|s\rangle)$ is informative about k-separability. In particular, a given state is \textit{genuinely} multipartite entangled if and only if $g (|s\rangle,{\bf v})$ is irreducible, that is $g (|s\rangle,{\bf v})$ is not a block-diagonal matrix for all possible chose of the unit vectors ${\bf v}^\mu$. This irreducibility condition indeed excludes k-separable cases, i.e. states $|s\rangle$ given as product of entangled states (e.g.  $|s\rangle = |s_1\rangle\otimes |s_2\rangle$).
From Eq. \eqref{gmunu} we derive
\begin{equation}
    \tr [ g(|s\rangle,{\bf v}) ]= \sum^{M-1}_{\mu=0} \left[1 - ({\bf v}^\mu\cdot \langle s |\bm{\sigma}^\mu|s\rangle )^2\right] \, ,
\end{equation}
that shows that the unit vectors
\begin{equation}
\tilde{\bf v}^\mu = \pm \langle s |\bm{\sigma}^\mu|s\rangle/\Vert \langle s |\bm{\sigma}^\mu|s\rangle \Vert \, ,
    \label{vtildes}
\end{equation}
provide the $\inf$ of $\tr(g)$.
Therefore, we obtain 
\begin{equation}
    E(|s\rangle) =M - \sum^{M-1}_{\mu =0} \Vert \langle s |\bm{\sigma}^\mu|s\rangle\Vert^2 \, .
    \label{measureMqpure}
\end{equation}
The $\inf$ operation, makes the measure \eqref{emeasure}
independent from the choice of the operators $U^\mu$. Consequently, its numerical
value is associated with the class of states generated by acting with local unitary
transformations on the state $|s\rangle$, and does not depend on a specific element
chosen inside the class. This is a necessary condition for a well defined
entanglement measure \cite{PhysRevLett.78.2275}.
The ED is built upon the geometrical structure of the projective Hilbert space through the principle of minimum distance expressed in Eq. \eqref{emeasure}.
The generalization of the ED to mixed states naturally leads to a measure of quantum correlations. We refer the reader to \cite{Adesso_2016} for a thorough definition of the various types of quantum correlations. In particular, the distinction between the set $\mathcal{S}$ of the separable states and the set $\mathcal{C}$ of the classical states (i.e. states with no quantum correlation) is of critical importance to understand how mixed states (unlike pure states) can contain quantum correlation even in the absence of entanglement.
In the following, we derive a quantum correlation measure for mixed states via the same minimum distance principle and, for this reason, we name it quantum correlation distance (QCD). Then, we derive an entanglement measure for mixed states from the QCD, through a regularization process of the density matrix.

\section{Entanglement Distance for Mixed States}

\subsection{\label{sec:qcd} Quantum Correlation Distance}

The Hilbert-Schmidt distance $D$ between two general square matrices, $A$ and $B$, is given by
\begin{equation}
D(A,B) = \sqrt{\dfrac{1}{2} \tr[(A-B)^\dagger (A-B)] } \, .
\label{DAB}
\end{equation}
We derive from the latter, the distance between two neighbouring density matrices, by
\begin{equation}
d^2_{_{dm}} (\rho,\rho+d\rho)=
\dfrac{1}{2} \tr[(d\rho)^\dagger (d\rho)]
\, .
\label{dm-distance}
\end{equation}
The infinitesimal variation $d\rho$ of state $\rho$ is
\begin{align}
d\rho &= \sum^{M-1}_{j=0} d\tilde{U}^\mu  \rho + \rho \sum^{M-1}_{\mu=0} d\tilde{U}^{\mu\dagger} \nonumber \\
&= -i \sum^{M-1}_{\mu = 0} \sum^3_{j=1} \comm{\sigma^\mu_j}{\rho}
n^\mu_j d \xi^\mu
\, ,
\label{drho}
\end{align}
where, the infinitesimal operators $d\tilde{U}^\mu$ have been defined in \eqref{dU} and with $\comm{}{}$, we mean the commutator.
We have
\begin{align}
d^2_{_{dm}} (\rho,\rho+d\rho)=
\sum^{M-1}_{\mu,\nu=0}g_{\mu\nu} (\rho, {\bf n})
 d\xi^\mu d\xi^\nu
\, ,
\label{dm-distance2}
\end{align}
where
\begin{equation}
\!\!\!\!\!\!  g_{\mu\nu} (\rho, {\bf n}) = \dfrac{1}{2}
\sum^3_{i,j=1} \tr [\rho \acomm{\sigma^\mu_i}{ \sigma^\nu_j} \rho -2 \rho \sigma^\mu_i
\rho\sigma^\nu_j ] n^\mu_i n^\nu_j \, ,
\label{gmunudm}
\end{equation}
with $\acomm{}{}$ we mean the anticommutator.
By imposing the minimum distance principle in a similar fashion to what we did above, we define the QCD for the state $\rho$ as
\begin{equation}
C(\rho) =\inf_{ \{{\bf n}^\nu\}_\nu} \tr (g(\rho, {\bf n}) ) \, .
\label{emeasurerho}
\end{equation}
We have
\begin{equation}
\sum^{M-1}_{\mu=0} g_{\mu \mu}(\rho, {\bf n}) = 
 M \tr(\rho^2)-\sum^{M-1}_{\mu=0} 
\sum^3_{i,j=1} \tr [\rho \sigma^\mu_i
\rho\sigma^\mu_j ] n^\mu_i n^\mu_j \,
\label{trg}
\end{equation}
Finally, by defining the matrices $A^\mu(\rho)$, for $\mu=0,\ldots,M-1$, whose entries are
\begin{equation}
A^\mu_{ij}(\rho) = \tr [\rho \sigma^\mu_i
\rho\sigma^\mu_j ] \,  ,
\label{Amu}
\end{equation}
we obtain the closed-form expression for the QCD of $\rho$,
\begin{equation}
C(\rho) 
=\sum^{M-1}_{\mu=0}\Big(\tr(\rho^2) -  \lambda^\mu_{max} (\rho) \Big) = \sum^{M-1}_{\mu=0}C_\mu(\rho)\, ,
\label{QCDrho}
\end{equation}
where, for $\mu=0,\ldots,M-1$, $\lambda^\mu_{max} (\rho)$ is the maximum of the eigenvalues of $A^\mu(\rho)$, and $C_\mu(\rho) = \tr(\rho^2) - \lambda^\mu_{max} (\rho)$ is the QCD of the subsystem $\mu$.
The QCD \eqref{QCDrho} fulfills the following requirements for a \textit{bona fide} measure of quantum correlation \cite{Adesso_2016}:
\begin{enumerate}
 \item \label{cond:null_for_classical}     $C_\mu(\rho)=0$ if  $\rho\in\mathcal{C}_\mu$, i.e. if $\rho$ is classical in the subsystem $\mu$. Indeed,  $\forall\rho\in\mathcal{C}_\mu$ we can write $\rho = \sum_j p_j\rho_j^{\mu_C} \otimes |j\rangle\langle j|^\mu $, where the $\{|j\rangle^\mu\}$ form an orthonormal basis in $\mathcal{H}^\mu$, $\mu_C$ is the complement of subsystem $\mu$ and $\sum_j p_j=1$. Then $\exists \bm{n}^\mu$ such that $\forall j,\, \bm{n}^\mu\cdot\bm{\sigma}^\mu|j\rangle^\mu=\pm1$, hence $\lambda^\mu_{max} (\rho)=1$ and $C_\mu(\rho)=0$. It results $C(\rho)=0$ if  $\rho\in\mathcal{C}$, i.e. if $\rho$ is fully classical.
    \item \label{cond:LU-invariant}$C(U\rho U^\dagger)=C(\rho)$, i.e. it is invariant under local unitary transformations.
    \item \label{cond:ent_for_pure}$C(\ket{\psi}\bra{\psi})=E(\ket{\psi})$, i.e. it reduces to the measure of entanglement on pure states defined by Eq. \eqref{measureMqpure}.
\end{enumerate}

\subsection{\label{sec:edm} Entanglement Distance}
As stated above, for a mixed state, the existence of quantum correlation is not a sufficient condition to guarantee the presence of entanglement. To extract from a given state $\rho$ its entanglement essence, we now propose a procedure of regularization of $\rho$, repurposing our measure of quantum correlations to catch the true degree of entanglement owned by $\rho$. 

Given a state $\rho$, we consider all of its possible decomposition $\{p_j,\rho_j\}$, such that
\begin{equation}
    \rho = \sum_j p_j \rho_j \, ,
\end{equation}
where $\sum_j p_j = 1$ and $\tr[\rho_j] = 1$. 
Also, we consider all the possible local partial transformation on qubit $\mu$:
\begin{equation}
    \rho_U^\mu (\{p_j,\rho_j,U_j^\mu\}) = \sum_j p_j U_j^\mu \rho_j U^{\mu\dagger}_j \, ,
    \label{rhoU}
\end{equation}
where, for each $j$, $U_j^\mu$ is an $SU(2)$ local unitary operator acting on qubit $\mu$. 
We define the ED for state $\rho$
\begin{equation}
E(\rho) = \inf_{\{p_j,\rho_j\}} \Big\{ \sum_{\mu=1}^{M-1} \inf_{\{U_j^\mu\}} C_\mu\left( \rho^\mu_U(\{p_j,\rho_j,U_j^\mu\})  \right) \Big\}\, .
\label{EDrho}
\end{equation}
Note that, similarly to the QCD, one can define $E_\mu(\rho)$ as the ED of subsystem $\mu$, simply discarding the complement in the sum on $\mu$ in \eqref{trg}. 
The ED \eqref{EDrho} fulfills the following requirements for a suitable measure of quantum
entanglement:

\begin{enumerate}[(i)]
    \item \label{cond:null_for_separable}
     $E_\mu(\rho)=0$ if $\rho\in\mathcal{S}_\mu$, that is if $\rho$ is separable in $\mu$. Indeed, it then admits a decomposition $\{p_j,\rho_j\}$, where, for each $j$, $\rho_j= (\mathbb{I}^\mu + \bm{\sigma}_{{\bf n}_j}^\mu)/2 \otimes \rho_j^{\mu_C}$, where. Thus, it is always possible to determine local partial operators $U_j^\mu$, such that, after transformation \eqref{rhoU} it results
    $\rho_U^\mu = \sum_j p_j|j\rangle\langle j|^\mu\otimes \rho_j^{\mu_C}$ and, from property \ref{cond:null_for_classical}, it follows our statement. It results $E(\rho)=0$ if $\rho\in\mathcal{S}$, that is if $\rho$ is fully separable.
    
    \item Reciprocally, if $E(\rho) =0$, then $\rho$ is separable. First of all, we note that, for each $\mu=0,\ldots,M-1$, $\lambda^\mu_{max} (\rho) \leq \tr(\rho^2)$. In fact, for each $\mu$ and for each unit vector ${\bf n}^\mu$ it is possible to determine a unitary local operator $U$, so that $\tr[\left(\rho (\bm{\sigma}_{{\bf n}})^\mu \rho (\bm{\sigma}_{{\bf n}})^\mu \right)] = \tr[\tilde{\rho} {\sigma}_3^\mu \tilde{\rho} {\sigma}_3^\mu]$, where $\tilde{\rho} = U\rho U^\dagger$. Furthermore $\tr[\tilde{\rho} {\sigma}_3^\mu \tilde{\rho} {\sigma}_3^\mu] = \sum_j \tilde{\rho}_{jj}^2+2 \sum_{i\neq j} \pm |\tilde{\rho}_{ij}|^2\leq \sum_j \tilde{\rho}_{jj}^2+2 \sum_{i\neq j}  |\tilde{\rho}_{ij}|^2 = \tr[\tilde{\rho}^2] = \tr[\rho^2]$. Moreover, for each pair $i\neq j$, $\exists\mu$ such that the term $|\tilde{\rho}_{ij}|^2$ appears in $\tr[(\tilde{\rho} {\sigma}_3^\mu)^2]$ with a negative sign. Yet, $E(\rho) =0$ implies that there exist a decomposition of $\rho$, let's say $\overline{\rho}$, for which 
    \begin{equation}
    \sup_{{\bf n}^\mu}  \tr[\overline{\rho} (\bm{\sigma}_{{\bf n}})^\mu) \overline{\rho} (\bm{\sigma}_{{\bf n}})^\mu)]  =\tr[\overline{\rho}^2] 
    \end{equation}
    for each $\mu$. We hence have $|\overline{\rho}_{ij}|^2 =0$ for each $i\neq j$. But this implies that $\overline{\rho}$ is diagonal and then $\rho$ separable.
    \end{enumerate}

For a given density matrix decomposition $\{p_j,\rho_j\}$, the minimization on the local unitary partial transformations, entailed by Eq. \eqref{EDrho}, can be addressed by studying the local minima of $C(\rho(\{p_j,\rho_j,U_j\}))$ under variation of $\{U_j\}$. 
Nevertheless, it can be proven that such fixed points do correspond only to cases where $E(\rho)=0$, hence to separable states. Therefore, the minima of \eqref{EDrho} in the case of non-separable states, do not correspond to fixed points, but rather to nonlocal 
(boundary) minima.
Remarkably, these fixed points of the minimization procedure \eqref{EDrho} can, at least in some cases, be realized by a decomposition $\{p_j,\rho_j\}$ including entangled pure states $\rho_j$. 
In particular, for two-qubits states diagonal in the Bell basis (the Bell-diagonal (BD) states, see \cite{PhysRevA.54.1838,PhysRevA.88.012120}) the fixed points can always be realized on the eigen-decomposition (hence, where the $\rho_j$ are Bell states).
This of course greatly simplify the problem, as the full exploration of the $\{p_j,\rho_j\}$-space is avoided.
It is worth emphasizing that BD states are representative of the larger class of two-qubits states of maximally mixed marginals (that is, for which $\forall\mu$ and $\forall j$, 
$\tr [\rho\sigma_j^\mu ]=0$ \cite{PhysRevA.54.1838}), hence \eqref{EDrho} is tractable in the same manner for this class of states.
Leaning on numerical evidences, we further conjecture that, for a given state $\rho(\pmb{\gamma})$ depending on parameters $\pmb{\gamma} = (\gamma_1,\gamma_2,...)$, the decomposition realizing the minimum \eqref{EDrho} is the same in the whole parametric domain of $\pmb{\gamma}$, and can hence be inferred from the fixed points found in the domains where this state is separable, if such a domain exists.
This suggests that the minimization over all possible decompositions $\{p_j,\rho_j\}$ might in fact possess \textit{non-trivial} general solutions \footnote{Here, by ``non-trivial solutions'' of the minimization procedure, we mean solutions which do not require to find the decomposition of $\rho$ in terms of pure product-states $\rho_j=\bigotimes_\mu (\mathbb{I}^\mu + (\bm{\sigma}_{{\bf n}_j})^\mu)/2$.}, depending on the considered class of states. A subsequent more thorough work on such a classification of the solutions of this procedure could thus lead to an entanglement measure of relatively low computational cost, in particular for systems symmetric under qudit permutations, and with low $\operatorname{rank}(\rho)$.

\section{Applications}

\subsection{Bell diagonal states}

As a first and seminal example of application of this procedure, we consider general BD states. They can be expressed as:
\begin{align}
\rho_{BD}(\{p_\alpha\}) &= \sum_{\alpha=1}^4 p_\alpha |\psi_\alpha\rangle\langle\psi_\alpha|\nonumber\\ &= \frac{1}{4}\Big(\mathbb{I} + \sum_i c_i \sigma_i^0\sigma_i^1 \Big)\,  ,
\label{rhoBD}
\end{align}
where the $|\psi_\alpha\rangle$ are the four Bell states: $|\psi_\pm\rangle=\frac{1}{\sqrt{2}}(|00\rangle \pm |11\rangle)$ and $|\phi_\pm\rangle=\frac{1}{\sqrt{2}}(|01\rangle \pm |10\rangle)$. Furthermore, we have $\forall i,\,|c_i|\leq1$, and the $c_i$ are such that the vector $(c_1,c_2,c_3)$, fully characterizing the state, belongs to the tetrahedron $\mathcal{T}$ of vertices $(-1,1,1),\,(1,-1,1),\,(1,1,-1),\,(-1,-1,-1)$. The separable BD states belong to the octahedron $\mathcal{O}$ of vertices $(\pm1,0,0),\,(0,\pm1,0),\,(0,0,\pm1)$, corresponding to the condition $\forall\alpha,\,p_\alpha\leq2$, and the classical BD states are located on the Cartesian axis $(c_1,0,0),\,(0,c_2,0),\,(0,0,c_3)$ \cite{PhysRevA.54.1838,PhysRevA.88.012120}.

\begin{figure}
    \centering
    \includegraphics[width=1.\linewidth]{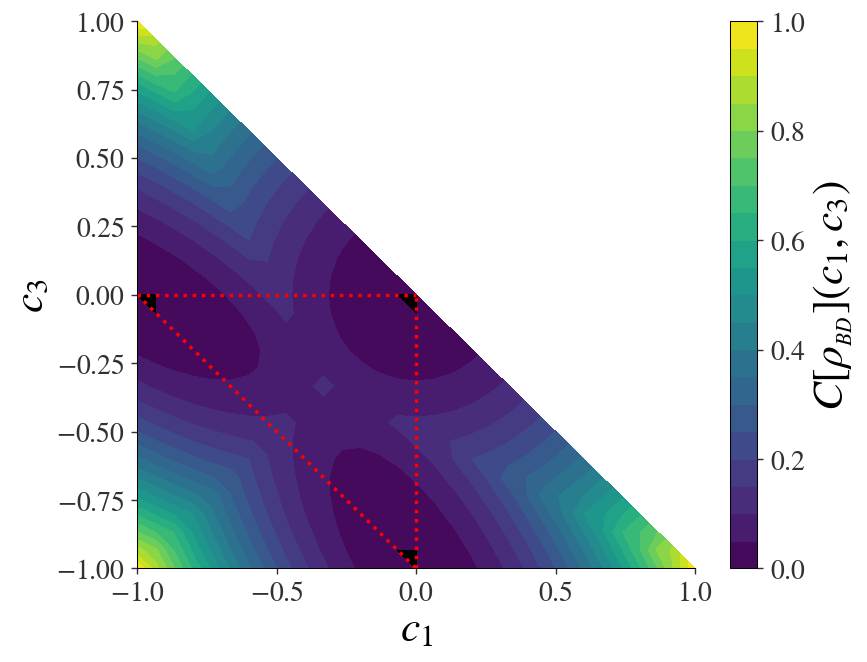}
    \caption{Quantum correlations $C[\rho_{BD}](c_1,c_2=c_1,c_3)/2$ for a face of the BD state tetrahedron $\mathcal{T}$, corresponding to a mixture of three Bell states. The red dotted line defines the smaller triangle where the state is separable, according to the PPT criterion \cite{PhysRevLett.77.1413,HORODECKI19961}. The vertices of the large triangle correspond to pure Bell states. Those of the red dotted triangle, of vanishing QCD, correspond to equal-weight mixtures of two Bell states, which are evidently the three only \textit{classical} states in the represented domain.}
    \label{figBD_QCD}
\end{figure}

\begin{figure}[ht!]
    \centering
    \includegraphics[width=1.\linewidth]{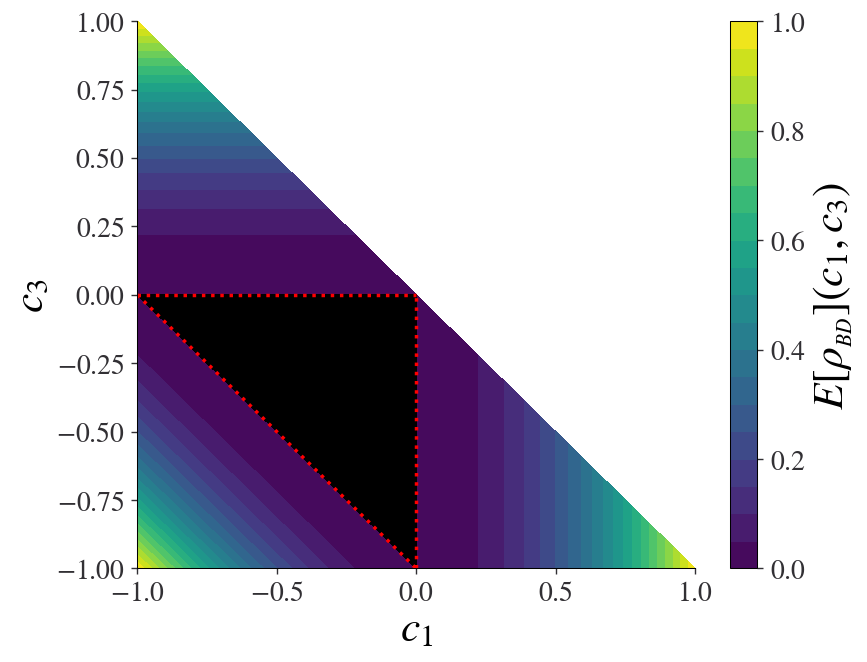}
    \caption{Entanglement distance $E[\rho_{BD}](c_1,c_2=c_1,c_3)/2$ for a face of the BD state tetrahedron $\mathcal{T}$, corresponding to a mixture of three Bell states. The red dotted line defines the smaller triangle where the state is separable, according to the PPT criterion \cite{PhysRevLett.77.1413,HORODECKI19961} and a number of alternative derivations available in the literature (see e.g. \citep{PhysRevA.54.1838}). Values below the threshold of $10^{-3}$ have been represented in black to emphasize that they correspond to a numerical zero, given the level of precision allowed by such time-costly minimization. The vertices of the large triangle correspond to pure Bell states, and those of the smaller black triangle to equal-weight mixtures of two Bell states.}
    \label{figBD_ED}
\end{figure}

Direct calculation yields the following result for the QCD of general BD states
\begin{equation}
    C(\rho_{BD}(\{p_\alpha\})=2\sum_{\alpha=1}^4 p_\alpha^2 -4\max_{P\{i,j,k,l\}}\Big\{p_ip_j +p_kp_l\Big\},
\end{equation}
where the maximum is taken on all permutations $P\{i,j,k,l\}$ of the indices $\{1,2,3,4\}$. Figure \ref{figBD_QCD} shows the QCD of BD states on a face of $\mathcal{T}$.
We were not able to find a simple analytic solution of the minimization procedure for the most general case of BD states. However, numerical minimization provided us with empirical evidence that the procedure (22) also leads for these states to the squared concurrence, as shown in figure \ref{figBD_ED}, which represent a face of the tetrahedral domain of BD states. It it interesting to note that the ED, as the concurrence and unlike the QCD, is constant on planes parallel to the boundary faces of the separability region: the ED of any given state indeed equates the QCD of the closest point located on a hinge of $\mathcal{T}$, hence the closest mixture of only two Bell states. 
\vskip0.4cm

\subsection{Werner states}

\begin{figure}
   \centering
    {\includegraphics[width=1.\linewidth]{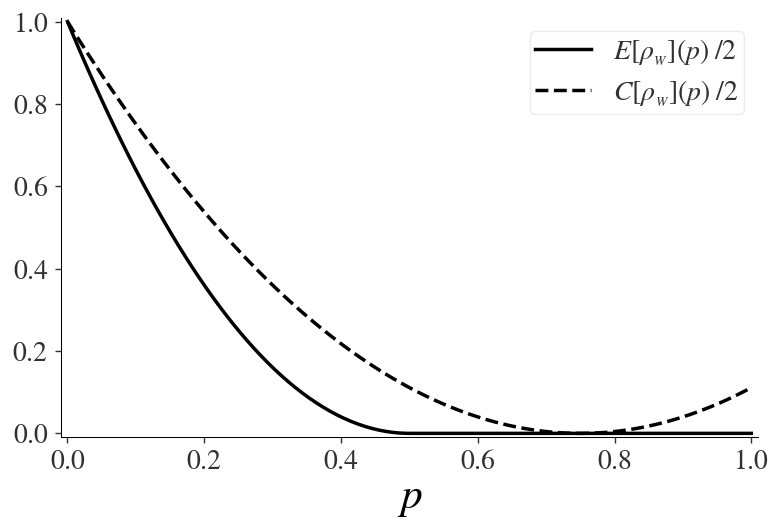}}
  \caption{$C[\rho_W](p) /2$ and $E[\rho_W](p) /2$ versus $p$ for state \eqref{rhoW}. It is clear that the state $\rho_W(p=0)$ is, as expected, the maximally-entangled, and that the states $\rho_W(p>1/2)$ are fully-separable, as can be verified using the PPT criterion \cite{PhysRevLett.77.1413,HORODECKI19961}. This plot emphasizes that separable states can contain quantum correlation (i.e. not be classical). Note that, here $E[\rho_W](p) /2=C_2^2[\rho_W](p)$, that is, the ED equates twice the squared concurrence for 2-qubits Werner states.}
      \label{figfirstfirst}
\end{figure}

Let us now consider the two-qubit Werner states (WS) \cite{PhysRevA.40.4277}, which stems as a special case of BD state, for which a simple analytical solution for the proposed procedure is available. WS are used as a testbed since they illustrate many features of mixed-states entanglement \cite{PhysRevA.54.3824}. Using Eq. \eqref{rhoBD}, they can simply be expressed as
\begin{equation}
\rho_{W}(p)=\rho_{BD}\big(\frac{p}{3},\frac{p}{3},\frac{p}{3},(1-p)\big) \, .
\label{rhoW}
\end{equation}
Via direct calculations, one gets for the QCD of the WS
\begin{equation}
C(\rho_W(p) ) = 2(1-\dfrac{4}{3} p)^2 \, .
\label{EWs}
\end{equation}
WS yields a relatively simple solution to the minimization procedure (22). Indeed, as it can be easily verified, if we set 
\begin{align}
    &U_{|\psi_+\rangle}(\theta)=U^\mu_z(\theta)U^\mu_x(\pi), \nonumber\\ &U_{|\psi_-\rangle}(\theta)=U^\mu_z(\pi-\theta)U^\mu_x(\pi), \text{ and}\nonumber\\
    &U_{|\phi_+\rangle}=U_{|\phi_-\rangle}=\mathbb{I},
\end{align}
with $\mu=0,1$ arbitrarily chosen, the fixed points are found for $\theta=\arccos{(\frac{3}{2p}-2)}$. This last expression has a solution if and only if $p\geq1/2$, which is the parametric region of separability for $\rho_W(p)$ (as can be verified by application of the positive partial trace criterion, see \cite{HORODECKI19961}). Hence, $E(\rho_W)=0$ for $p\geq1/2$. 
For $p<1/2$ numerical minimization yields $E(\rho_W)=4p^2 - 4p + 1$. This corresponds to $\theta=0$ uniformly on this whole domain, which is also the value previously determined at $p=1/2$: hence, the minimum after this point cease to be a fixed point, but keeps the last position in terms of the parameters governing the rotations. One can understand this as the fixed point reaching the boundary of the parametric domain as the geometry of the state is changing continuously, becoming a simple point on a slope, located at this boundary.
All together, for Werner states, the result of our entanglement measure exactly equates twice the square of the concurrence \cite{PhysRevLett.80.2245}, that is
\begin{equation}
E(\rho_W(p) ) = 2\Theta(1/2-p)(1-2p)^2\, ,
\label{EDWs}
\end{equation}
Fig. \ref{figfirstfirst} shows $C(\rho_W(p) ) /2$ versus $p$, there it is clear that the only state with no quantum correlation, i.e. \textit{classical state} according to the conventional terminology \citep{Adesso_2016}, is the one corresponding to the value $p=3/4$, whereas the maximally quantum-correlated state is that of $p=0$. On the other hand, the state is entangled only in the region $p<1/2$, and separable otherwise, a well-known fact that can be easily checked by application of the positive partial transpose (PPT) criterion \cite{PhysRevLett.77.1413,HORODECKI19961}. Alternatively, one can find, in the separable region, the expression of $\rho_W$ convex combination of (non-orthogonal) product states, using a more involved calculation resorting to the so-called Bloch representation.. \\

\section{Summary}
Our goal in this work has been to derive an entanglement measure from geometric properties of the projective Hilbert space describing a quantum multipartite system. We have derived from a minimum distance principle an explicitly computable entanglement measure for pure states and a quantum correlation measure either for pure or mixed states. We have shown that the quantum correlation measure reduces to the entanglement one, in the case of pure states. 
For mixed states, the quantum correlation is not a faithful measure of entanglement. To extract from a given state $\rho$ its entanglement essence, we have defined a regularization procedure for the density matrix that allows our measure of quantum correlation to catch the true degree of entanglement owned by $\rho$. We have proved that the entanglement and quantum-correlation measures derived satisfy the requirements for suitable measures of these quantities.
Finally, we have applied our procedure to two relevant examples, the Bell diagonal states and the Werner states.

\begin{acknowledgments}
We acknowledge support by the QuantERA ERA-NET Co-fund 731473 (Project Q-CLOCKS), in addition, we acknowledge the support by National Group of Mathematical Physics (GNFM-INdAM). We also acknowledge discussions with Lorenzo Capra.
\end{acknowledgments}

\appendix

\nocite{*}
\bibliography{references}

\end{document}